\documentclass[twoside,12pt]{article}
\usepackage{epsfig}

\def\Journal#1#2#3#4{{#1} {#2} (#4) #3 }

\def\PRC{{\em Phys. Rev.} C}

\newcommand{\be}{\begin{equation}}
\newcommand{\ee}{\end{equation}}
\newcommand{\bea}{\begin{eqnarray}}
\newcommand{\eea}{\end{eqnarray}}

\begin{document}

\title{ \vspace{1cm} Understanding Supernova Neutrino Physics using Low-Energy Beta-Beams}
\author{N.\ Jachowicz$^{1}$ and G.C.\ McLaughlin$^2$ \\ 
\\
$^1$Ghent University, Department of Subatomic and Radiation Physics,\\ Proeftuinstraat 86, B-9000 Gent, Belgium \\
$^2$Department of Physics, North Carolina State University, Raleigh, \\North Carolina 27695-8202}
\maketitle

\begin{abstract}
We show that fitting linear combinations of low-energy beta-beam spectra to supernova-neutrino energy-distributions reconstructs the response of a nuclear target to a supernova flux  in a very accurate way.  This allows one to make direct predictions about the supernova-neutrino signal in a terrestrial neutrino detector.
\end{abstract}

As the main messengers reaching a terrestrial detector from a future supernova explosion, neutrinos are  the principal candidates for shedding light on the supernova puzzle.  However, the interpretation of the information they are able to provide is limited by our understanding of the signal in a detector.  Nuclei are an important target material, as they offer relatively large cross sections with thresholds in the energy range of importance. Unfortunately, there is only little experimental data available, and theoretical calculations tend to be subject to model-dependent uncertainties.   

We propose  constructing synthetic spectra as linear combinations of low-energy beta-beam spectra~\cite{vol}
\begin{equation}
n_{fit}({\varepsilon})=\sum_{i=1}^N a^{\gamma_i} n^{\gamma_i}(\varepsilon),
\label{const}
\end{equation}
where the values of gamma, and the coefficients $a^{\gamma}$, are determined by
 minimizing the expression
\begin{equation}
\int d\varepsilon \, |n_{fit}(\varepsilon)-n_{SN}(\varepsilon)|.
\label{fit}
\end{equation}
In this way, we obtain constructed spectra that are as close as possible to the original supernova-neutrino distribution \cite{keil}.
We repeated the procedure with N=3 and N=5 beta-beam spectra in the fit, restricting the choice for $\gamma$ to integer values between 5 and 15.
The folded cross sections obtained using the constructed spectra are in very good agreement with those resulting from a folding procedure with  the original energy distributions.

As the energy transfer to the target nucleus determines the excited state the nucleus will be left in, and hence the reaction products that will be observed in the detector, it is the central quantity to study. 
Fig.~1 shows that the agreement between the folded cross sections is very good even for small numbers of beta-beam spectra in the fit, and for a broad range of average energies and widths of the energy distribution.

\begin{figure*}[tb]
\vspace*{14.cm}
\special{hscale=35 vscale=35 hsize=1500 vsize=600
         hoffset=-15 voffset=370 angle=-90 psfile="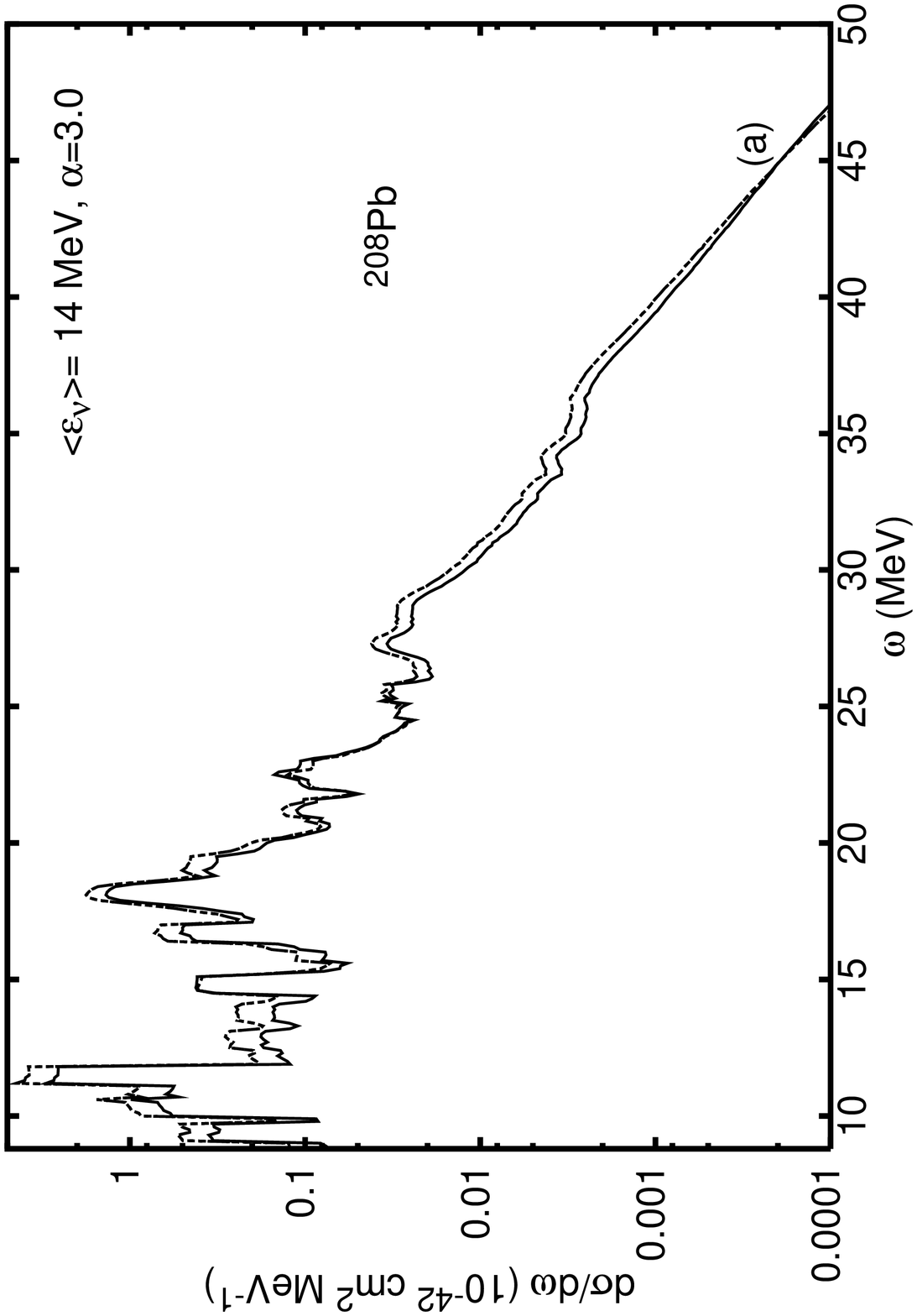"}
\special{hscale=35 vscale=35 hsize=1500 vsize=600
         hoffset=235 voffset=370 angle=-90 psfile="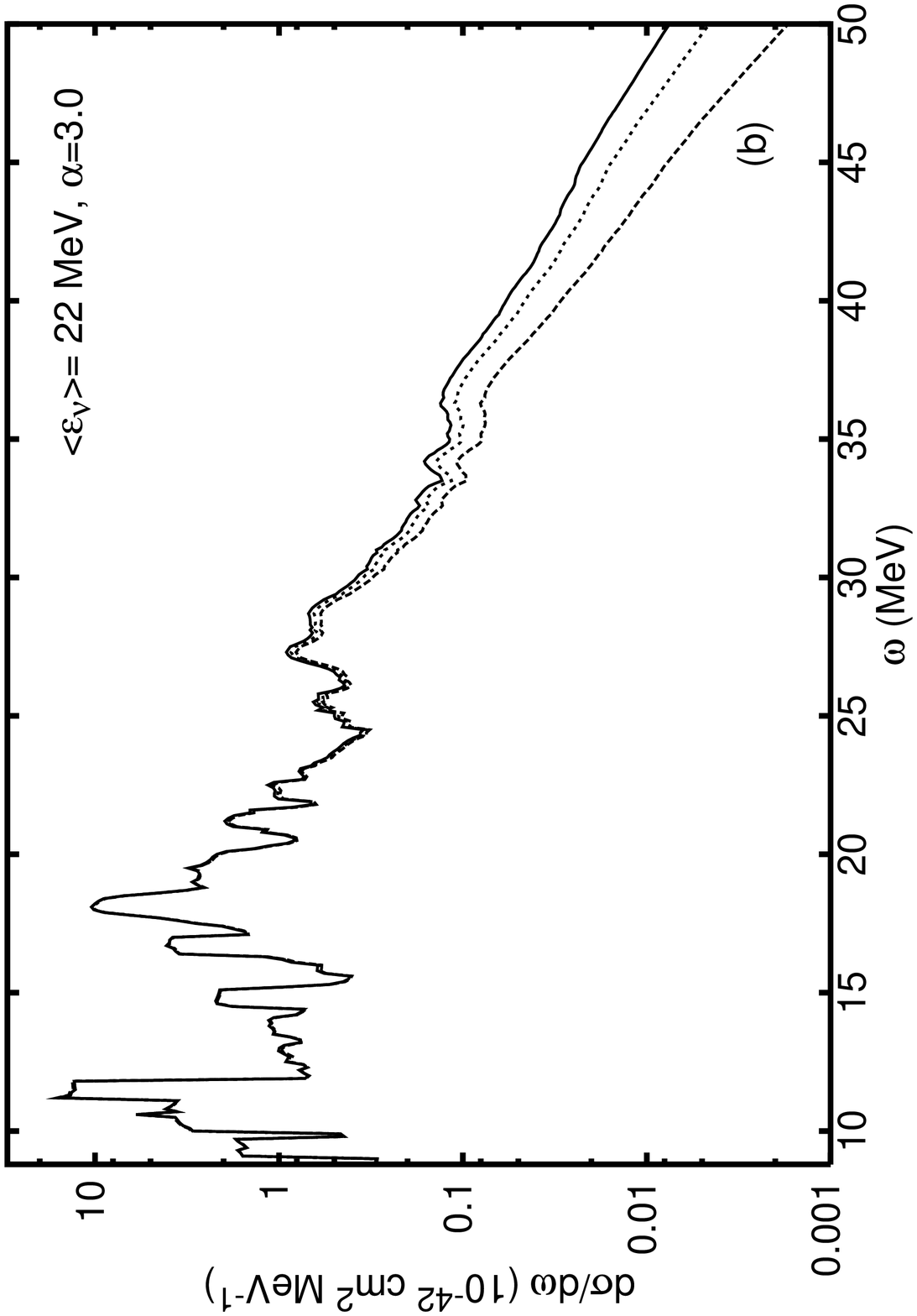"}
\special{hscale=35 vscale=35 hsize=1500 vsize=600
         hoffset=-15 voffset=195 angle=-90 psfile="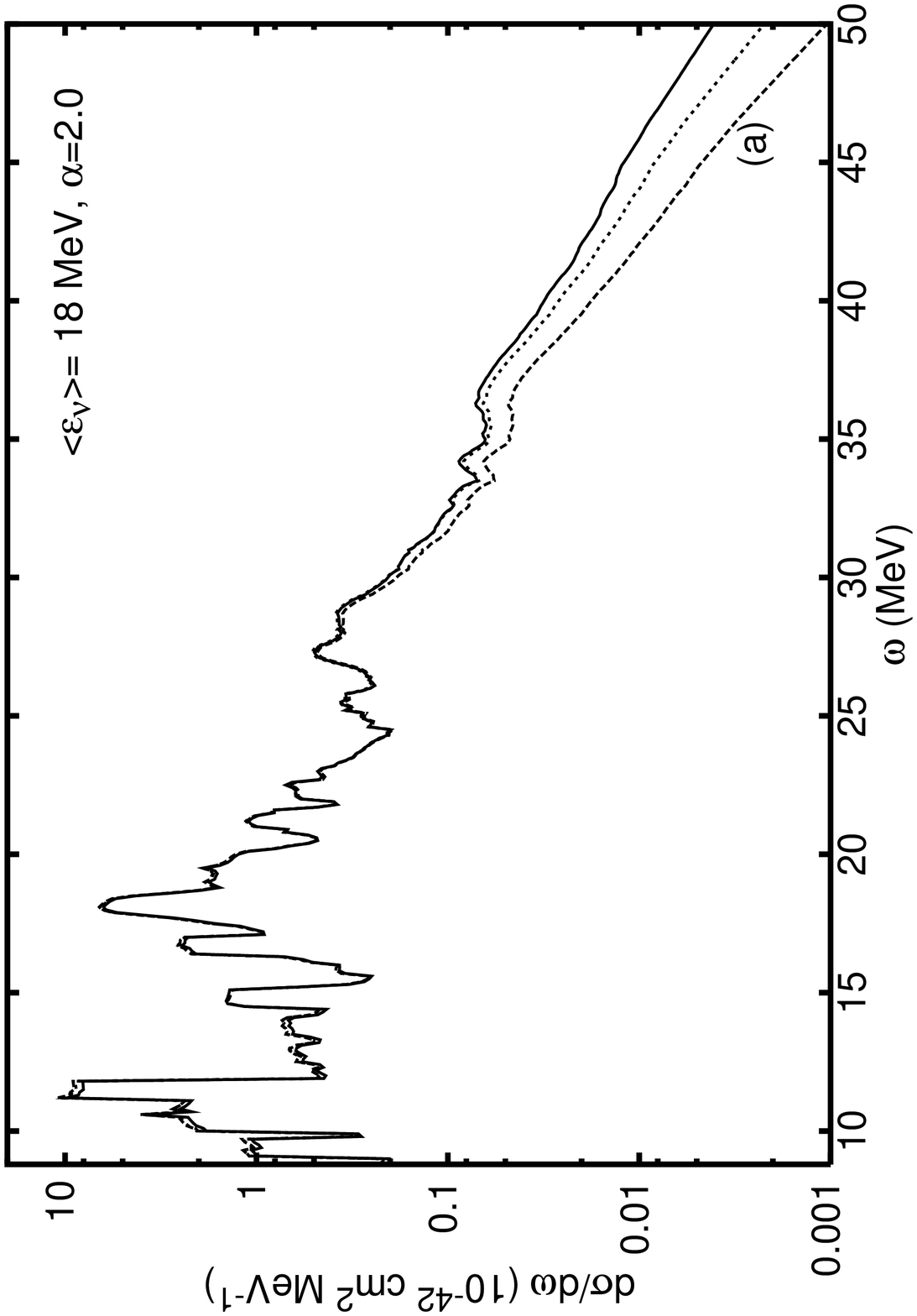"}
\special{hscale=35 vscale=35 hsize=1500 vsize=600
         hoffset=235 voffset=195 angle=-90 psfile="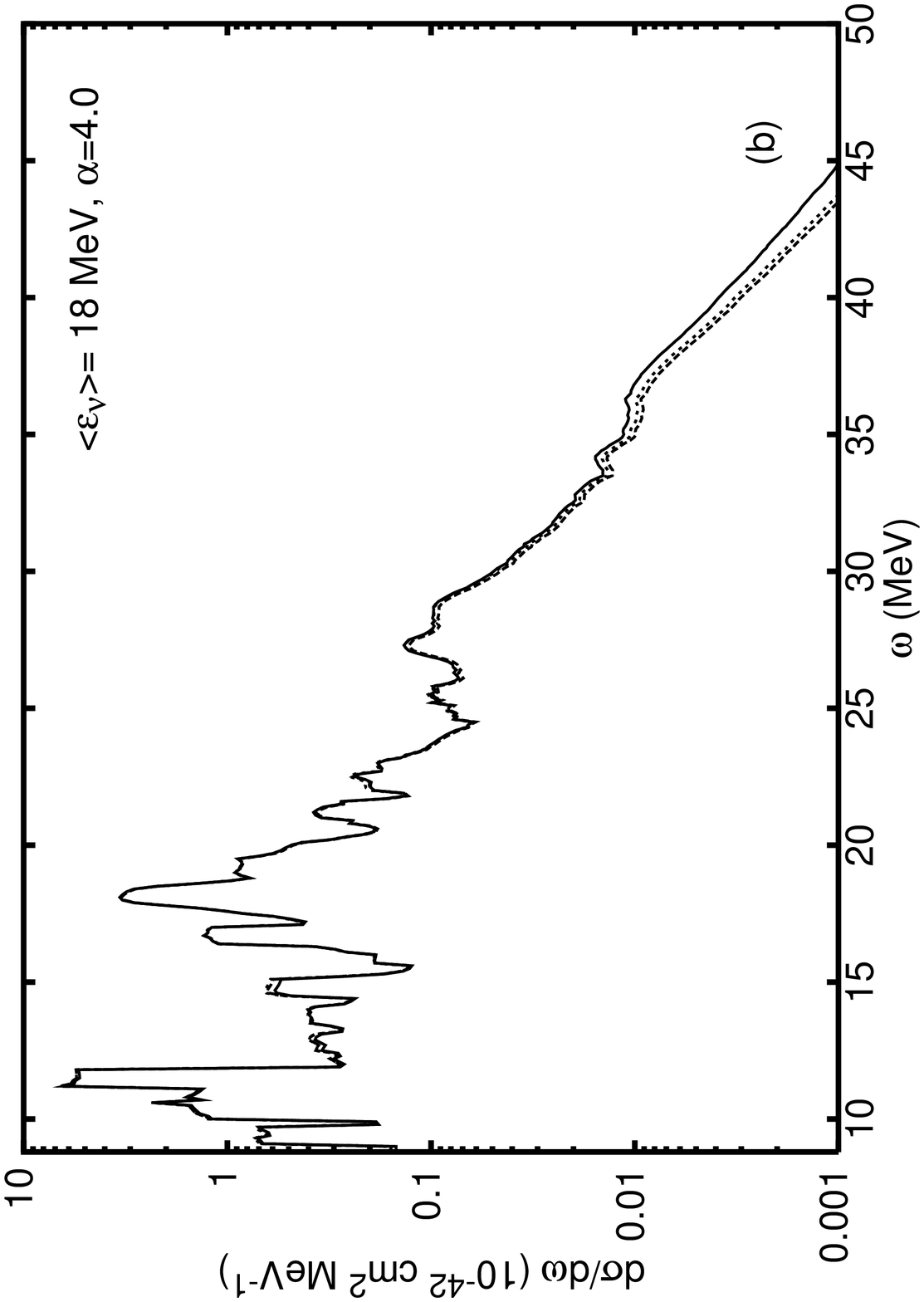"}
\caption{Comparison between differential cross section for neutral-current scattering on $^{208}$Pb, folded with a power-law supernova-neutrino spectrum (full line) and synthetic spectra with 3 (dashed line) and 5 components (full line) for different power law energy distributions : $\langle\varepsilon\rangle$=14 MeV, $\alpha$=3 (a), $\langle\varepsilon\rangle$=22 MeV, $\alpha$=3 (b), $\langle\varepsilon\rangle$=18 MeV, $\alpha$=2 (c), and $\langle\varepsilon\rangle$=18 MeV, $\alpha$=4 (d).}
\label{pbdif}
\end{figure*} 

This method allows one to  predict the supernova-neutrino response directly, starting from a number of beta-beam measurements, but without having to rely on theoretical cross-section predictions.
Reversing the fit, the technique  provides a way to reconstruct the main parameters of supernova-neutrino energy-distributions  from the supernova-signal in a terrestrial detector \cite{us}.
Analysis of the method shows that it is able to distinguish between spectra roughly 1 MeV apart in average energy \cite{us}.  Moreover the reconstruction is stable against  experimentally-inferred variations in the nuclear cross-sections  \cite{us}.



\end{document}